# Characteristics of Transposable Element Exonization within Human and Mouse

**Noa Sela[1¤], Britta Mersch[2], Agnes Hotz-Wagenblatt[2*], Gil Ast[1*]**

**1** Department of Human Molecular Genetics, Sackler Faculty of Medicine, Tel Aviv University, Tel Aviv, Israel, **2** Department of Molecular Biophysics, German Cancer Research Center (DKFZ), Heidelberg, Germany

## Abstract

Insertion of transposed elements within mammalian genes is thought to be an important contributor to mammalian evolution and speciation. Insertion of transposed elements into introns can lead to their activation as alternatively spliced cassette exons, an event called exonization. Elucidation of the evolutionary constraints that have shaped fixation of transposed elements within human and mouse protein coding genes and subsequent exonization is important for understanding of how the exonization process has affected transcriptome and proteome complexities. Here we show that exonization of transposed elements is biased towards the beginning of the coding sequence in both human and mouse genes. Analysis of single nucleotide polymorphisms (SNPs) revealed that exonization of transposed elements can be population-specific, implying that exonizations may enhance divergence and lead to speciation. SNP density analysis revealed differences between *Alu* and other transposed elements. Finally, we identified cases of primate-specific *Alu* elements that depend on RNA editing for their exonization. These results shed light on TE fixation and the exonization process within human and mouse genes.

**Funding:** This work was supported by a grant from the Israel Science Foundation (ISF 61/09), Joint Germany-Israeli Research Program (ca-139), Deutsche-Israel Project (DIP MI-1317), and Israel Cancer Research Foundation (ICRF). B.M. and A.H.-W. were supported by the Joint Germany-Israeli Research Program (ca-139) (DKFZ/MOST). N.S. is supported by the LMUexcellence fellowship. The funders had no role in study design, data collection and analysis, decision to publish, or preparation of the manuscript.

**Competing Interests:** The authors have declared that no competing interests exist.

* E-mail: gilast@post.tau.ac.il (GA); hotz-wagenblatt@dkfz-heidelberg.de (AHW)

¤ Current address: Department of Botanic, Faculty of Biology, Ludwig-Maximilians-University, Munich, Germany

## Introduction

The draft sequences of the human and mouse genomes confirmed that transposed elements (TEs) have played a major role in shaping mammalian genomes [1,2]. Sequences of transposed elements comprise at least 45% of the human and 37% of the mouse genomes (Lander et al., 2001; Waterston et al., 2002). A large fraction of the TEs were inserted into transcribed regions, mostly within intronic sequences [3]. These intronic insertions contributed to the enlargement of intron size within mammalian genomes (Lander et al., 2001; Waterston et al., 2002). Sironi et al. identified constraints on insertion of TEs within introns [4] and showed that gene function and expression influence insertion and fixation of distinct transposon families in mammalian introns [5].

Exonization is the creation of a new exon as a result of mutations in intronic sequences [6], whereas intronization is the creation of a new intron. TEs have enriched the human transcriptome by exonizations [7] and intronizations [3]. In human, most of the exons that originated from TEs are from the primate-specific transposon called *Alu*. *Alu* elements are the most abundant repetitive elements in the human genome; there are upwards of 1.1 million copies, accounting for more than 10% of the human genome [1,8]. *Alu* elements are derived from the 7SL RNA [9]. The major burst of *Alu* retroposition took place 50–60 million years ago and has since dropped to a frequency of one new retroposition for every 20–125 births [10,11]. *Alu*-mediated mutagenesis, mostly through nucleotide insertions, has been estimated to be involved in close to 1% of Mendelian genetic disorders [12]. The occurrence of single nucleotide polymorphisms (SNPs) in and around *Alu* sequences has been discussed [8,13].

Makalowski and coworkers were the first to describe *Alu* elements within mature mRNA in human [14]. It is now clear that transposed elements are found within a large number of mature mRNAs [15]. The new exons generated from *Alu* elements are usually alternatively spliced; these exons comprise ~5% of alternatively spliced exons in the human transcriptome [16]. Exonized TEs that are alternatively spliced are not unique to human as most of the exonized TEs in the mouse genome are also alternatively spliced [3]. The molecular mechanism leading to *Alu* exonization has been well characterized. A typical *Alu* is around 300 nt and contains two similar monomer segments joined by an A-rich linker and a poly(A) tail-like region. *Alu*s insert into introns of primate genes by retrotransposition, usually in the antisense orientation. Eighty-five percent of exonizations have occurred from the right arm in the antisense orientation [3,16]. The poly(A) tract of this arm in the antisense orientation creates a strong polypyrimidine tract (PPT). Downstream from this PPT a 3′ splice site is selected and further downstream from that site (approximately 120 nt) a 5′ splice site is recognized [17]. Without the left arm, exonization of the right arm shifts from alternative to constitutive splicing. This results in elimination of the evolutionary conserved isoform and may thus be selected against [18]. Only one or two mutations are required within intronic *Alu*s that reside in antisense orientation relative to the coding sequences to yield a consensus 3′ splice site [19] or 5′ splice site [20]. The role of splicing regulatory sequences on the exonization process has also been studied [21,22,23]. The 3′ splice site of exonized *Alu*s are very similar to those of the 3′ splice sites of mammalian interspersed repeat (MIR) exons [24].

Recent studies indicate that the pattern of splicing of exonized TEs differs among human tissues [25,26,27]. Additionally, there are variations in splicing patterns within individuals in the human population [28,29,30]. Certain SNPs correlate with heritable changes in alternative splicing but do not cause disease, thus indicating a link between genetic variation and mode of splicing [29,31,32]. Another study identified SNPs correlated with obesity that cause variation within alternative splicing patterns [28].

The exonization process is subject to many evolutionary constraints: New exons are generally alternatively spliced [7] and the inclusion rate is relatively low [16,19,33]. This implies that novelties added to established genes (within established coding sequences, CDSs) are under lower purifying selection if they do not interfere with the original coding sequence, compared to those events that change the original CDS. Also, exonization usually occurs in untranslated regions (UTRs) [3] or within duplicated genes [34], further supporting the idea that purifying selections are more intense on exonization events that occur within CDSs. Thus, alternative splicing of *Alu* exons enriches the human transcriptome with new mRNAs without eliminating the original, functionally important transcripts, which are generated via exon skipping [35].

Here we set to find additional characteristics of TE exonization events within human and mouse. We looked at the location of the exonizations within genes and the SNP densities, and evaluated SNPs that change canonical splice sites. We found that exonizations occur preferentially in the beginning of protein coding sequences. Moreover, we show that exonizations can be population specific. Our findings reveal a possible contribution of TE exonizations to population divergence within human and mouse.

## Results

### The locations of TE exonizations within coding sequences

Non-symmetrical, conserved, alternatively spliced exons are more often located at the beginning of the CDS than elsewhere in transcripts [36,37,38]. We analyzed the Transpogene database of exons that originated from TEs [39] to determine whether there is a bias in their location within mRNA. We normalized the CDS length between 0 and 1 (see Materials and Methods) and compared, in increments of 0.1, the extent of TE exonization at different locations in human and mouse (Figure 1). We found that exonized TE sequences are biased to reside in the first half of the CDS sequence compared to alternatively spliced cassette exons that did not originate from TE exonizations. Most exonizations in both human and mouse are found between position 0.1 and position 0.4 within the CDS, with a median location of 0.336 in human and 0.369 in mouse. No statistically significant differences were observed between the human and mouse populations or within different TEs families. Alternatively spliced cassette exons that did not originate from TEs are found at a median location of 0.513 and 0.507 in human and mouse, respectively. Statistically significant differences were observed between alternative cassette exons and TE exons (Wilcoxon Rank Sum test, $p = 1.2244e{-}027$ and $p = 1.2322e{-}006$ for human and mouse, respectively). These results imply that most TE exonizations tend to occur within the first introns of genes. In human non-TE alternatively spliced exons, 1353 out of 17,642 are the second exon, whereas in TE-derived exons 233 out of 927 are found in the first intron and if spiced become the second exon; this difference is statistically significant (Fisher's exact test, $p < 10^{-42}$). The first intron is substantially longer, with respect to the other introns, in most human and mouse genes and shows higher rate of TE insertion [39]. The longer introns presumably provide a good environment for exonization [40]. Effects of TE exonization within the first intron are usually neutral with respect to the protein sequence, but can affect signal sequences [41].

In order to analyze whether the location bias results from potential involvement of purifying selection, we separated our data to three groups: exonizations that contain an in-frame stop codon (599 exons), exonizations that are non-symmetrical and do not contain an in-frame stop codon (216 exons), and symmetrical exons that do not contain stop codons (137 exons). The median locations within the normalized CDS of these three groups are 0.3062, 0.3795, and 0.4199, respectively. The Wilcoxon Rank Sum test showed that there is a statistically significant difference between the first and the third group ($p = 0.0428$) but not between the second group and the third group or the first group and the second ($p = 0.2555$ and $p = 0.3641$, respectively). This observation strengthens the hypothesis that the 5′ position bias of TE exonization has a connection with the NMD machinery. We previously showed that non-symmetrical exons (not related to TEs) that are alternatively spliced in both human and mouse (and thus likely to be functional events) tend to be located near the 5′ end of the CDS, whereas conserved symmetrical alternative exons are located throughout the CDS [37]. The current results show a statistically significant difference in location between symmetrical exons and those with in-frame stop codons. We hypothesize that TE-driven alternative exons are under purifying selection to be locate at the beginning of the CDS, presumably to enhance identification of the TE-containing mRNA by the nonsense-mediated decay (NMD) system [42].

### SNP density within intronic and exonized TEs

Identifying features shaping the architecture of sequence variations is important for understanding genome evolution and mapping of disease loci. A positive correlation was shown previously between *Alu* elements and SNPs density [13]. Analysis of the positive association between schizophrenia and a cluster of SNPs and haplotypes in the seventh intron of the β2 subunit of the type A γ-aminobutyric acid receptor revealed that the *Alu*-Y near the 5′ end of exon 8 contains as many 11 SNPs [43].

Here we set out to evaluate and compare SNP densities in all TE families from human and mouse. All positions of exons and introns of all genes as annotated in the Golden Path database and the positions of intergenic regions along with the number of SNPs in these regions were obtained and divided by the total length of the particular region. The dataset contained 39,288 human genes. For the human analysis of the SNPs, we evaluated 382,892 exons with 446,357 SNPs, 347,948 introns with 8,428,718 SNPs, and 8,899 intergenic regions with 10,395,717 SNPs. We also used 31863 mouse genes. For the mouse analysis we evaluated 301506 exons with 273700 SNPs, 270782 introns with 500541 SNPs, 8602 intergenic regions with 661474 SNPs.

Multiplying the resulting SNP densities by 100 yielded the SNP frequency per 100 bp. The average SNP density in the human genome is 0.43 in exons, 0.4 in introns, and 0.41 in intergenic regions. The similar densities of SNPs in exons, introns, and intergenic sequences were somewhat unexpected, as one might expect strong evolutionary pressure against substitutions in protein coding regions. This might be caused by a bias of the SNP data from dbSNP itself as EST data is the basis for many SNPs. In the mouse genome, the average frequency of SNPs is 0.31, 0.33, and 0.28 in exons, introns, and intergenic regions, respectively. These SNP densities are consistent with the number of SNPs observed in the baseline windows presented in Figure 2 for human TEs and in Figure 3 for mouse TEs. These results are in agreement with the SNP densities previously obtained from exons, introns, and intergenic regions in human and mouse genomic sequences [13].

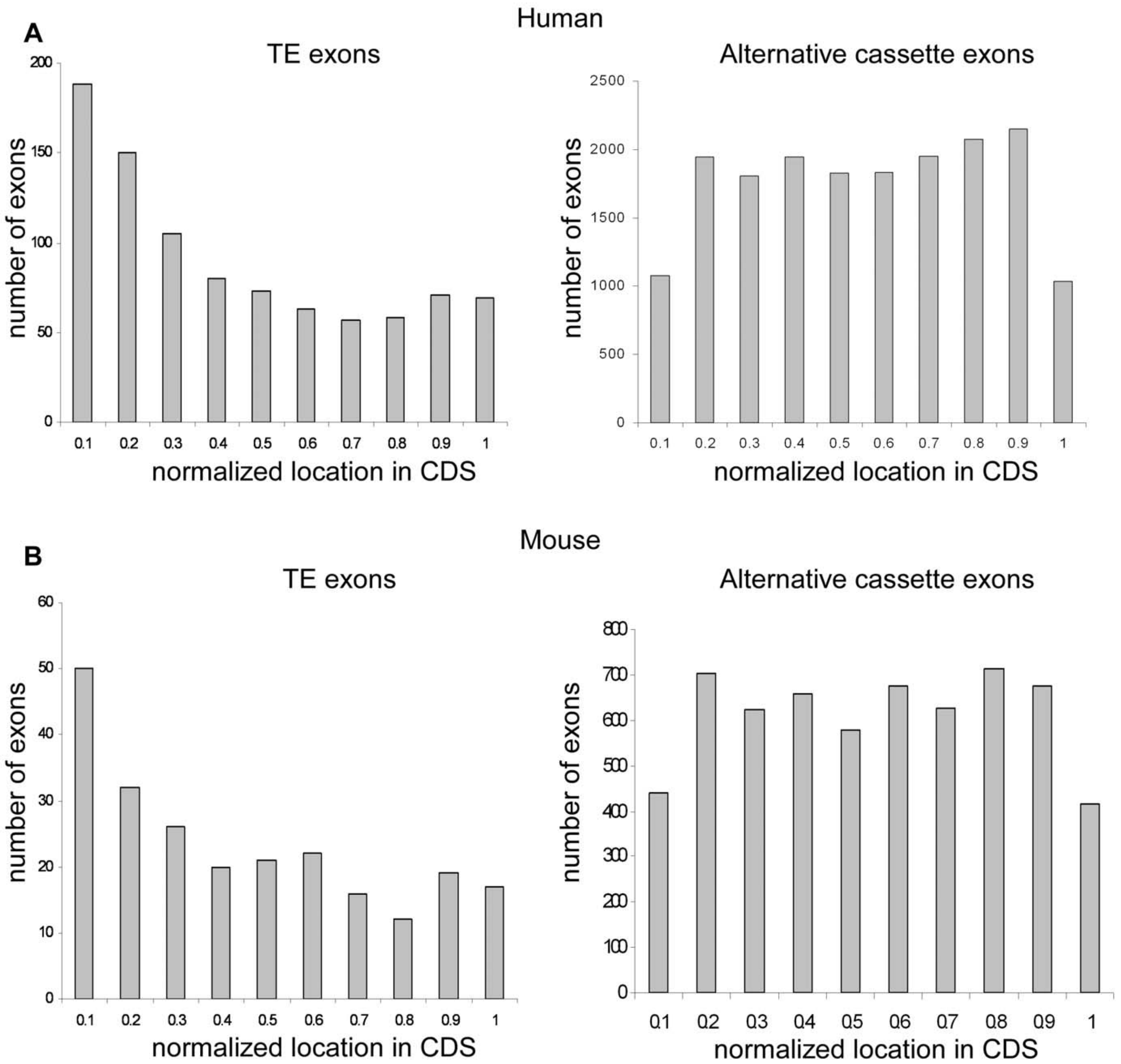

**Figure 1. Bias toward exonization at the 5′ end of the CDS.** TE-derived exons (left panels) and alternatively spliced cassette exons that did not originated from TEs (right panels) are shown in normalized locations along the CDS in increments of 0.1 (exon locations were normalized between 0 and 1, see Materials and Methods) for (**A**) human and (**B**) mouse. The x-axis is the normalized CDS location and the y-axis is the number of alternative exons.
doi:10.1371/journal.pone.0010907.g001

As shown in Figure 2, the SNP density in primate-specific *Alu* elements is 0.53, which is higher than the baseline level. The density in *Alu* elements is the highest level observed among the different families of TEs. *Alu* elements are GC rich with 24 or more CpG dinucleotides per element. These dinucleotides are prone to mutation as a result of deamination of 5-methylcytosine. Only half of the SNPs in young *Alu* elements were found at CpG dinucleotides, however [8,20,44]. Also, analysis of the GC-rich *Alu* body separately from the AT rich *Alu* tail showed that both parts are enriched in SNPs [13]. Therefore, the GC content cannot be the sole determinant of this enrichment. For the L1 elements, the SNP density is similar to the baseline frequency, whereas the frequency is lower than baseline for the other families of TEs. A correlation of the age of the different *Alu* families with the SNP density shown by Ng et al. [13] suggests that the lower SNP density for L1 and the other TE elements might be related to their earlier integration into the human genome. However, we cannot rule out the option that there is not a simple correlation between the age of the TE and the number of SNPs. The primate-specific *Alu* element and the rodent-specific B1 element originated from the same 7SL RNA gene and share a high level of sequence identity. Nevertheless, the high SNP density detected in *Alu* elements was not observed in murine B1 elements (Figure 3).

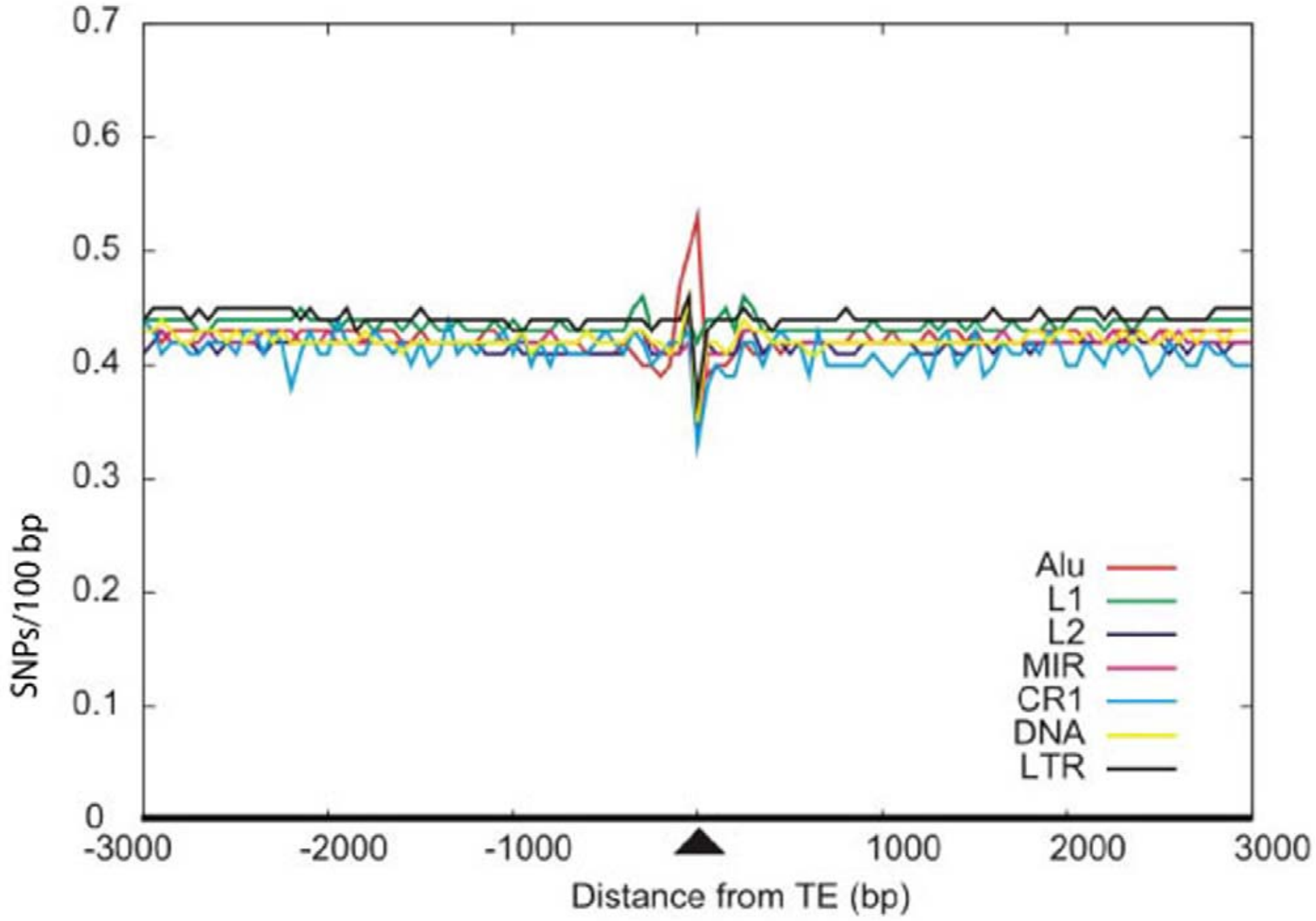

**Figure 2. Density of SNPs within all transposed elements in the human genome.** The average SNP frequency in the TE-body and the flanking sequences is shown in a sliding window of 50 bp. All frequencies are normalized to a frequency per 100 bp. The center of the TE is located at position 0.
doi:10.1371/journal.pone.0010907.g002

We then examined the SNP density in exonized TEs (Table 1). The SNP density in exonized TEs from all TE families in the human genome is lower than the overall SNP density of all TEs, but the difference is not significant (Mann-Whitney test, $p = 0.382$, two-tailed). An exception was observed in the CR1 (LINE-3) elements; exonized CR1 elements have a higher than average SNP density. However, only four CR1 elements were exonized so the sample size is very small. In mouse, for all transposed element families, the density of SNPs in exonized TEs was significantly higher than the overall density in all TEs (Mann-Whitney test, $p = 0.004$, two-tailed). In mouse, exonization seems to occur preferentially in areas with higher SNP density.

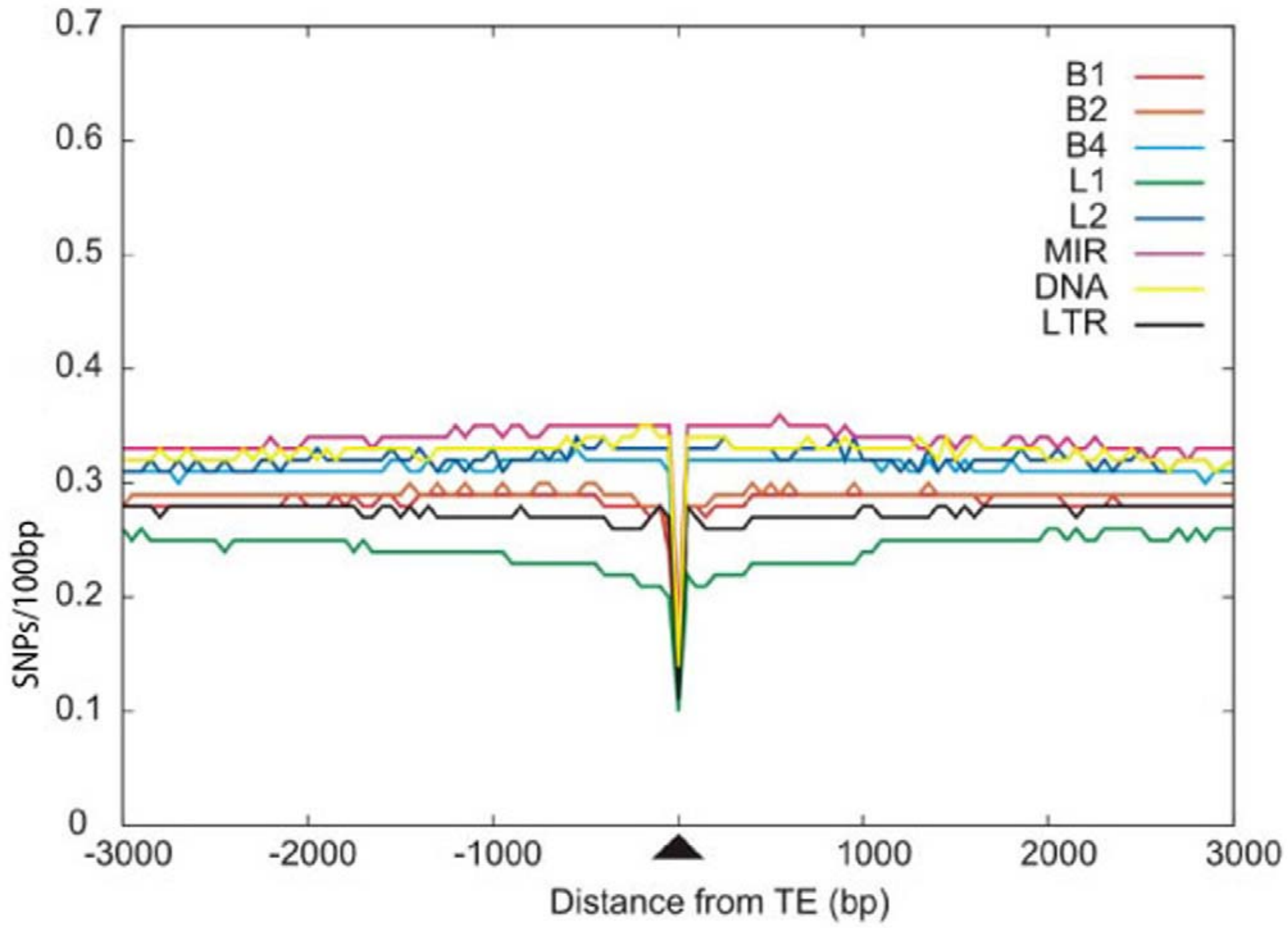

**Figure 3. Density of SNPs within all transposed elements in the mouse genome.** The average SNP frequency in the TE and the flanking sequence is shown in a sliding window of 50 bp. All frequencies are normalized to a frequency per 100 bp. The center of the TE is located at position 0.
doi:10.1371/journal.pone.0010907.g003

**Table 1.** Densities of SNPs in exonized TEs and all TEs in the human and in the mouse genomes.

| Human | | | Mouse | | |
|---|---|---|---|---|---|
| TE family | SNP density exonized | SNP density all | TE family | SNP density exonized | SNP density all |
| *Alu* | 0.45 | 0.53 | B1 | 0.29 | 0.12 |
| L1 | 0.37 | 0.42 | B2 | 0.27 | 0.12 |
| L2 | 0.33 | 0.34 | B4 | 0.16 | 0.14 |
| MIR | 0.28 | 0.33 | L1 | 0.15 | 0.10 |
| CR1 | 0.51 | 0.33 | L2 | 0.21 | 0.16 |
| LTR | 0.31 | 0.37 | MIR | 0.30 | 0.17 |
| DNA | 0.23 | 0.35 | LTR | 0.25 | 0.11 |
| | | | DNA | 0.0 | 0.14 |

## SNPs in the splice sites of exonized TEs may cause variation in the exonization process

In order to investigate the possibility that exonization of TEs creates transcriptomic diversity within the human population, we searched for SNPs that eliminate or create canonical splice site in a TE. Specifically, we looked either for changes in the invariant AG dinucleotide at the 3′ splice site or the canonical GT or GC at the 5′ splice site. Although there are other positions that might alter recognition by the splicing machinery, only the four positions must be fully conserved to ensure selection by the spliceosome. To enhance the fraction of bona fide exonization events we searched for exonized TEs that are supported by at least two ESTs. Our analysis revealed 10 SNPs in canonical splice sites of TE-derived exons in the human genome (Table 2); these SNPs eliminate change a canonical splice site into a non-canonical one (the ancestral nucleotides are also shown in Table 2). Of the ten, five are in the acceptor and five in the donor splice sites. Seven of the SNPs occur in splice sites of exonized *Alu* elements, two in splice sites of exonized L2 elements, and one in the splice site of an exonized LTR element. To ensure that we identified the sequence without the SNP correctly, we examined the sequences of the orthologous TEs in chimp (Table 2). Additional support for the role of SNPs in TE population-specific exonization is given by the ssSNPTarget database (http://sssnptarget.org/) [45], the SNPs rs2377301 and rs5758111 have EST evidence for exon skipping due the SNP modification. In the mouse genome, three splice sites of exonized TEs contain SNPs (Table 3). SNPs were found in the splice sites of an exonized B1 element, an exonized B2 element, and an exonized LTR element; all are within 5′ splice sites.

**Table 2.** SNPs in splice sites of exonized TEs in the human genome.

| Gene id | Chr./strand | Start–end | TE family | SNP info[1] | Position | Sequence in other species |
|---|---|---|---|---|---|---|
| RCSD1 | chr1/+ | 164341465–607 | *Alu* | rs1890128 (A/G) | 1st pos. donor | Chimp–G<br>Rhesus-G |
| FAM35A | chr10/+ | 88900743–863 | *Alu* | rs3129523 (A/T) | 2nd pos. donor | Chimp–T<br>Rhesus-A |
| TSFM | chr12/+ | 56463664–702 | *Alu* | rs2014886 (A/G) | 2nd pos. donor | Chimp–G<br>Rhesus-G |
| ETFA | chr15/− | 74389327–460 | *Alu* | rs2469213 (C/T) | 1st pos. acc. | Chimp–C<br>Rhesus-C |
| DPP9 | chr19/− | 4670214–336 | *Alu* | rs3059236 (-/TTTA) | 2nd pos. acc. | new insertion no chimp/rhesus info. |
| ZNF544 | chr19/+ | 63440426–512 | *Alu* | rs12979599 (A/G) | 1st pos. acc. | Chimp–G<br>Rhesus–no *Alu* insertion |
| LOC63929 | chr22/+ | 39581181–297 | *Alu* | rs5758111 (A/G) | 2nd pos. acc. | Chimp–A<br>Rhesus-A |
| ACTG2 | chr2/+ | 74041405–549 | L2 | **rs1721244 (A/G)** | 1st pos. donor | Chimp–G<br>Rhesus-G |
| CANT1 | chr17/− | 74505824–963 | L2 | **rs2377301 (C/T)** | 1st pos. acc. | Chimp–C<br>Rhesus–C<br>Mouse-C |
| AK129982 | chr8/+ | 12346635–774 | LTR | rs1988623 (A/G) | 1st pos. donor | Chimp–T<br>Rhesus–no *Alu* insertion |

[1]SNPs with population specific data are in bold.

We searched the NCBI Database of Single Nucleotide Polymorphisms for population frequency data. Data were only available for two of the 10 SNPs observed in the human genome (Table 4). One of them, SNP rs1721244, is located at chr2 position 73983403 and is the first nucleotide of the 5′ splice site. The allele with G has a canonical splice site (GT) but the other allele has a non-canonical splice site (AT). Both splice sites occur at a frequency of more than 0.3 (Table 4); thus, this SNP, and associated splice variation, is common in the human population. In this analysis, we selected only cases in which SNPs clearly changed the sequence directly at the splice site. We did not take into account SNPs within other splice signals or within exonic or intronic splicing enhancers/silencers that might modulate the selection level of the exon. Thus, the effect of SNPs on splicing might be greater than observed here.

We have also built a dataset of TEs with non-canonical splice sites that appear to be active based on evidence of exonization from ESTs or cDNAs. We searched the SNP database for SNPs that might change the non-canonical splice sites into canonical ones. In the human genome, we found 45 SNPs that changed a non-canonical splice site into a canonical site (a GT/GC dinucleotide in the 5′ splice site and an AG dinucleotide in the 3′ splice site; see supplementary data Table S1). Only three such SNPs were identified in the mouse genome. As a result of these SNPS, these exons are flanked by canonical acceptor and donor splice sites, explaining their identification by the splicing machinery and their presence in the ESTs database.

Population frequency data were available for 11 of the 45 SNPs (see supplementary data Table S2). One interesting case is SNP rs231518 in an L1 element. There are six ESTs and cDNAs with the 5′ splice site sequence AT, but the SNP rs231518 has a canonical 5′ splice site GT. The two alleles have an intriguing evolutionary history. There is a G at the 5′ splice site in chimp and orangutan and an A in rhesus. The sequences of chimp, orangutan, and rhesus were extracted from published sequences and the multi-species alignment of the SNP location was downloaded from UCSC genome browser [46]. We cannot exclude the possibility that A/G polymorphisms also exist within chimp, orangutan, and rhesus based on available data. The SNP rs231518 with the canonical dinucleotide 5′ splice site GT is the most frequent allele in all human populations (G allele frequency of 0.792 in the CEU population, 1 in the HCB and JPT population and 0.937 in the YRI population, see supplementary data Table S2).

### TE exons that depend on editing for their exonization

How new exons are created and established is an intriguing issue. Recently, Lev-Maor et al. [47] demonstrated that exonization of an *Alu* exon in the NARF gene depends on an RNA editing mechanism. In this case, editing from AA to AI activated the 3′ splice site; inosine is recognized as G by the splicing machinery [48]. We searched for additional cases in which the 3′ splice site of the exonized *Alu*s is AA or the 5′ splice site is AT, such that RNA editing to AG or GT, respectively, would produce a canonical splice site. We did not find any evidence for editing in 5′ splice sites of *Alu*-derived exons. However, we found six cases of *Alu* exonization in which the 3′ splice site contains an AA at the genomic level and EST sequences support exonization (Table 5). Two of these cases were found in ESTs generated from brain tissues and another two were from immune system tissues, tissues that have high levels of RNA editing [49,50,51,52]. Two other cases were found in cancerous tissues and in kidney. The most convincing evidence of exonization of an *Alu* element resulting from RNA editing is found within a non-coding brain-specific gene NR_024561. This exonization is supported by a validated Refseq sequence and three additional cDNA and ESTs (all from brain tissues). Moreover, transcripts containing this exon have three additional A-to-I editing sites within the *Alu*-derived exon. Several potential editing sites are usually observed within a region that contains two *Alu* elements located in opposite orientation due to the formation of a long double-stranded RNA structure between the elements [52]. Interestingly, the nearest *Alu* to that exonized in the NR_024561 gene is in the downstream intron (Figure 4). There is an *Alu* within the upstream intron but it is more than >2000 nucleotides away and is therefore unlikely to hybridize with the *Alu* exon [49,50,51,52]. NR_024561 appears to be a non-coding gene and is expressed exclusively in the brain. A BLAST search against the database of known non-coding RNAs NONCODE [53,54] revealed 85% identity (E value = 4e−52) of the NR_024562 isoform to the MESTIT1 non-coding RNA [55]. This isoform also had 86% identity (E-value = 6e–48) to the brain-specific non-coding KLHL1 antisense RNA [56]; this RNA is involved in the spinocerebellar ataxia type 8 (SCA8) neurodegenerative disorder [57,58].

## Discussion

### TE-derived exons are most often located near the 5′ end of the CDS

Cassette exons that are non-symmetrical and conserved in both human and mouse are more often located in the 5′ region of the coding sequence than in other regions [37]. Inclusion of non-symmetrical exons is likely to cause a frame shift in the coding sequence, introducing a premature stop codon and activating nonsense mediation decay or producing an unstable protein [36,38,59]. Most TE-derived exons are non-symmetrical [3,16] and are usually exonized from the first introns of a coding gene. We previously suggested that the majority of the TE-derived exons are non-symmetrical because they are still young in evolutionary terms and thus have not yet undergone purifying selection, which eliminates deleterious exonizations. Given a sufficient period of time, some of the currently non-symmetrical exons that are only mildly deleterious will eventually become symmetrical (through small deletions/insertions) and thus will add coding capacity into already established genes. Examples of functional TE-exonizations are exon 8 of ADAR2 gene [60] and exon 8 of NARF gene [47]. Nonsense codons in the 3′ halves of genes may less efficiently activate the RNA degradation machinery than those found near the start of a transcript [42,61]; it may also be that longer peptides are more likely to be deleterious than shorter ones [36,37,38]. The first intron is usually longer than the others and thus following exonization the two flanking introns are still relatively long. Alternatively spliced exons are generally flanked by longer introns than are constitutively spliced exons [62]. It is also possible that the bias observed may be due to the fact that TEs are more often found near the start of genes than in other regions. These results suggest that the first intron with its longer size function of a "buffer zone" to the emergence of new potentially deleterious exons.

**Table 3.** SNPs in splice sites of exonized TEs in the mouse genome.

| Gene id | Chr./strand | Start–end | TE family | SNP info | Position |
|---|---|---|---|---|---|
| Csrp2bp | chr2/+ | 143828541–730 | B2 | rs29540199 (C/T) | 2nd pos. donor |
| Zfp644 | chr5/− | 105752526–733 | B1 | rs33626312 (C/T) | 1st pos. donor |
| Rbm6 | chr9/− | 107929610–717 | LTR | rs33287617 (C/T) | 2nd pos. donor |

**Table 4.** Population frequency data for the human SNPs which occurred in the splice sites of the exonized TEs.

| | | Genotype detail | | | Alleles | |
|---|---|---|---|---|---|---|
| **SNP id** | **Population[1]** | **A/A** | **A/G** | **G/G** | **A** | **G** |
| rs1721244 (A/G, donor 1st position) | CEU | 0.27 | 0.57 | 0.17 | 0.55 | 0.45 |
| | HCB | 0.31 | 0.5 | 0.19 | 0.56 | 0.44 |
| | JPT | 0.1 | 0.46 | 0.44 | 0.33 | 0.67 |
| | YRI | 0.21 | 0.56 | 0.23 | 0.49 | 0.51 |
| | | **C/C** | **C/T** | **T/T** | **C** | **T** |
| rs2377301 (C/T), acceptor 1st positon) | CEU | 0.76 | 0.21 | 0.03 | 0.875 | 0.15 |
| | HCB | 0.44 | 0.45 | 0.11 | 0.67 | 0.33 |
| | JPT | 0.71 | 0.22 | 0.07 | 0.82 | 0.18 |
| | YRI | 0.8 | 0.2 | 0.0 | 0.9 | 0.1 |

[1]CEU–European, HCB–Asian, JPT–Asian, YRI–Sub-Saharan African.

## SNP densities vary depending on TE families

*Alu* elements were inserted into the human genome after the insertion of other families, such as MIRs, DNA transposed elements, and LTRs [64]. *Alu* elements show higher level of exonization than all other TE families [3]. Here we show that Alu elements tend to accumulate more SNPs than other TE families. The higher mutation rate in *Alu* elements is not correlated with their CpG enrichment [13,63]. There appears to be a correlation between the age of TE transposition and the mutation rate. A small fraction of L1 elements are still active in the human genome [64] and on average L1 elements contain a higher density of mutations than other analyzed families (L2, MIR, DNA, LTR). The average SNP density in TEs in the mouse genome is lower than the SNP density in the surrounding sequences. The SNP density in TEs in the human genome is at least 2-fold higher than that in mouse TEs. Artificial selection and inbreeding accompa-

**Table 5.** *Alu* exons edited at 3′ss.

| # | *Alu* type | Exon coordinates[1] | Gene | ESTs/cDNA accessions confirming the editing | Location of the closest intronic *Alu* | Other editing sites within the exon |
|---|---|---|---|---|---|---|
| 1 | *Alu*Jb in sense | chr1:52,768,028–52,768,145 | ZCCHC11-zinc finger, CCHC domain containing 11 isoform | BU178489–retinoblastoma | Upstream *Alu*Jb in antisense | No other editing sites |
| 2 | *Alu*Jb in antisense | chr5:61,653,166–61,653,305 | KIF2A–Homo sapiens kinesin heavy chain member 2A | AA834569-germinal center b cell tissue | Downstream *Alu*Sx in sense | One another editing site within the exon |
| 3 | *Alu*Jo in sense | chr6:24,489,146-24,489,281 | DCDC2-doublecortin domain containing 2 | BP332729-renal proximal tubule | Upstream *Alu*Sx in antisense | One another editing site within the exon |
| 4 | *Alu*Sx in sense | -chr16:36,579-36,721 | POLR3K-DNA directed RNA polymerase III polypeptide K | CR994793–t-lymphocytes | Upstream *Alu*Sg in antisense | No other editing sites |
| 5 | *Alu*Jo in sense | chr17:37,652,211-37,652,327 | STAT5B-Homo sapiens signal transducer and activator of transcription 5B | DA223574–brain | Downstream *Alu*Sg in antisense | Two more editing sites within the exon |
| 6 | *Alu*Jo in sense | chr19:40,262,281-40,262,395 | LOC100128675 -Homo sapiens hypothetical LOC100128675 non-coding RNA | NR_024561<br>AK124779<br>DA216531<br>DA216526–all from the brain | Downstream *Alu*Jb in antisense | Three more editing sites within the exon |

[1]Based on version hg18 of the human genome.

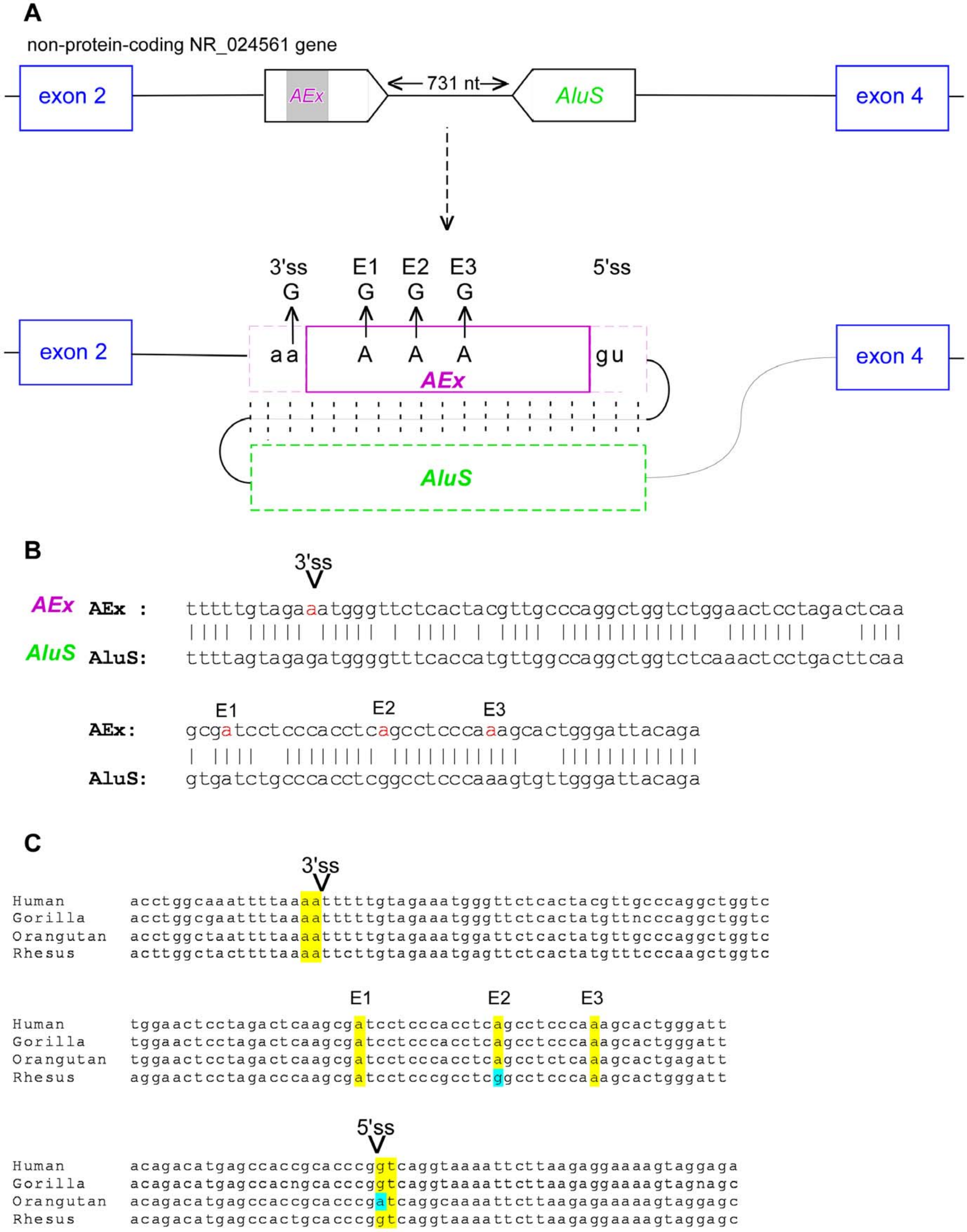

**Figure 4. Exonization of *Alu* element in NR_024561 dependent on RNA editing.** Editing was inferred from alignment of cDNAs to human genomic DNA. (**A**) Schematic illustration of exons 2 to 4 of the non-coding gene NR_024561. Exons are depicted as blue boxes. The *Alu*-exon, derived from *Alu*Jo (marked AEx; shown by purple box), is in an antisense orientation and is shown in the middle. The intronic, sense-orientation *Alu* sequence

(*Alu*S) is 731 base-pairs downstream of the exonized *Alu*. Sense and antisense *Alu*s are expected to form double-stranded RNA, thus allowing RNA editing. RNA editing changes an AA dinucleotide into a functional AG 3′ splice site (lower panel). RNA editing also occurs in three positions in the *Alu*-derived exon (E1, E2, and E3). (**B**) Predicted folding of the sense and antisense *Alu* sequences (upper and lower lines, respectively). Adenosines that undergo editing are marked by red. Splice sites utilized for *Alu* exonization are marked as 5′ss and 3′ss on the alignment. (**C**) Alignment of this region from four species: human, gorilla, orangutan, and rhesus. The 5′ splice site, 3′ splice site, and the three editing positions are marked in yellow.

nying the generation of laboratory mouse strains presumably serves to reduce genomic differences between individual mice. Therefore SNP data from mouse probably do not reflect real population dynamics.

### *Alu* exonization is coupled to the RNA editing mechanism

In our analysis, we found evidence for exonization of an *Alu* element that probably requires RNA editing. The NR_024561 gene is expressed exclusively in the brain. The exonized *Alu* element is from *Alu*Jo subfamily and it was inserted into this gene about 25 million years ago [65]. The 5′ splice site dinucleotide GT is conserved in rhesus and gorilla but not in orangutan. The 3′ splice site dinucleotide AA and the editing sites E1 and E3 are conserved in rhesus, orangutan, and gorilla (Figure 4C). The editing site E2 is not conserved in rhesus but is found in orangutan and gorilla. The conservation of these editing sites implies a possible function for this *Alu* exonization in this non-coding, brain-specific gene.

In summary, exonization of regions of transposed elements is thought to be an important contributor to mammalian evolution and speciation. We found that exonization of transposed elements is biased towards the beginning of the coding sequence in both human and mouse genes. Analysis of SNPs revealed population-specific exonization events, implying that exonizations may enhance divergence. These results shed light on TE fixation and the exonization process within human and mouse genes.

## Materials and Methods

### Dataset of TE exonizations within human and mouse protein coding genes

The dataset of human and mouse transposed element exonization was obtained from the TranspoGene database [39]. Based on UCSC genome browser annotations [66] of the human genome version hg17 and mouse genome version mm6. Sequences of TE exonizations within human and mouse protein coding genes were selected.

### Normalization of exon location

Exon location was determined by using the knownGene table downloaded from the UCSC genome browser. In this table, all genes are listed along with their CDS start and end coordinates. To normalize the exon location within the CDS, we calculated the location for the start point of the exon in the CDS without exceeding the boundaries of the CDS (N = CDS length − exon length + 1). The normalized location was the quotient of the actual location of the exon start point within the CDS divided by N.

### Cassette exon dataset

In order to create a dataset of cassette exons that had not originated from TE exonization, we downloaded the altSplice table from the UCSC genome browser [46,67]. We analyzed only the cassette exons dataset. We used GALAXY [68] and RepeatMasker in order to extract the sequences and exclude cassette exons that originated from TEs [69,70,71,72].

### SNP density in the TE families

SNP locations (original from dbSNP, http://www.ncbi.nlm.nih.gov/projects/SNP/) were obtained from the UCSC Genome Browser Database [66] (versions hg17, May 2004 for human and mm6, March 2005 for mouse). For every family of TEs the average SNP density in the TE-body was determined. For comparison purposes, the SNP density in sequences surrounding the TEs was extracted in 50-bp non-overlapping windows from either end of the TE up to a distance of 3 kb. This yielded 120 windows which we call *baselines*. The positions of all TEs in the genome and locations of SNPs within each TE were determined using the SNP data set from UCSC Genome Browser Database. The same was done for the surrounding 50-bp non-overlapping windows (up to distance of 3 kb) for determination of the baseline density of SNPs. The SNP densities were averaged over all TEs and normalized to SNP frequency per 100 bp by dividing the average number of SNPs within the TE by the average length of the TEs divided by 100. Averaging the SNP frequencies in all 50-bp windows flanking the TE yielded the baseline SNP frequency, similar to the calculation described in [13]. The number of SNPs in each of the 50-bp windows was multiplied by 2 to obtain the frequency per 100 bp. The SNP density in exonized TEs was then determined. Exons originating from exonizations of TEs that were flanked by canonical splice sites and that had at least two ESTs confirming their exonization were used. The average SNP density in the exonized TEs was determined for the human and mouse. All SNP densities are the SNPs per 100 bp.

### SNPs in the splice sites of the exonized TEs

Annotations of SNPs were obtained from the UCSC Genome Browser Database [66] (versions hg17, May 2004 for human and mm6, March 2006 for mouse). A search for SNPs in splice site dinucleotides of exonized TEs was conducted. Any changes from GT or GC dinucleotides in the first two positions of the intron (5′SS) and AG dinucleotides in the last two positions of the intron (3′SS) by SNPs were considered; these mutations change a canonical splice site into a non-canonical one thus eliminating the selection of this exon by the splicing machinery. We also considered situations in which SNPs changed a non-canonical splice site into a canonical one if at least one transcript confirmed the existence as exon.

Population frequency data was obtained from the NCBI Database of Single Nucleotide Polymorphisms (dbSNP Build ID: 125) [73]. This data was only available for a small number of SNPs in dbSNP. Many researchers do not provide genotype or frequency data in their submissions. dbSNP Build ID 125 had approximately 27 million SNPs and only 3.5 million of these had frequency data associated with them.

### Dataset of *Alu* exonization resulting from editing of the 3′ splice site

The dataset of *Alu* exonizations was searched for *Alu* elements with the non-canonical AA 3′ splice sites or the AT non-canonical 5′ splice site. These *Alu*s were filtered according to the following criteria: (1) no SNPs were detected within these slice sites, (2) at least one A to G transition was detected between the DNA sequence and the mRNA, and (3) another *Alu* sequence in reverse orientation is located within a distance of 2000 bp.

## Supporting Information

**Table S1** SNPs in non-canonical splice sites of exonized transposed elements in the human genome as well as in the mouse genome resulting in a canonical splice site. Given are the gene id, the chromosome and strand on which the SNP is located, the start and end of the exon which derived from the transposed element, the transposed element's family, the SNP id and the alleles of the SNP and the position at which the SNP is located (always seen from the exon, that is, 1st position of acceptor indicates the base which is located nearest to the splice site).
Found at: doi:10.1371/journal.pone.0010907.s001 (0.05 MB DOC)

**Table S2** Population frequency data for the SNPs which changed a non-canonical splice site into a canonical one while the other splice site was already canonical. Given is the SNP id along with the alleles and the position where this SNP occurred as well as the frequency data. Here, the homozygosity for the first allele, the heterozygosity, the homozygosity for the second allele, the Hardy-Weinberg proportions as well as the frequencies for each of the alleles are given. CEPH-European, HISP-Hispanic, AD-African American, CEU-European, HCB-Asian, JPT-Asian, YRI-Sub-Saharan African, HWP-Hardy-Weinberg proportions.
Found at: doi:10.1371/journal.pone.0010907.s002 (0.10 MB DOC)

## Author Contributions

Conceived and designed the experiments: NS GA. Performed the experiments: NS. Analyzed the data: NS BM AHW. Wrote the paper: NS BM AHW GA.

## References

1. Lander ES, Linton LM, Birren B, Nusbaum C, Zody MC, et al. (2001) Initial sequencing and analysis of the human genome. Nature 409: 860–921.
2. Waterston RH, Lindblad-Toh K, Birney E, Rogers J, Abril JF, et al. (2002) Initial sequencing and comparative analysis of the mouse genome. Nature 420: 520–562.
3. Sela N, Mersch B, Gal-Mark N, Lev-Maor G, Hotz-Wagenblatt A, et al. (2007) Comparative analysis of transposed element insertion within human and mouse genomes reveals Alu's unique role in shaping the human transcriptome. Genome Biol 8: R127.
4. Sironi M, Menozzi G, Comi GP, Bresolin N, Cagliani R, et al. (2005) Fixation of conserved sequences shapes human intron size and influences transposon-insertion dynamics. Trends Genet 21: 484–488.
5. Sironi M, Menozzi G, Comi GP, Cereda M, Cagliani R, et al. (2006) Gene function and expression level influence the insertion/fixation dynamics of distinct transposon families in mammalian introns. Genome Biol 7: R120.
6. Kister KP, Eckert WA (1987) Characterization of an authentic intermediate in the self-splicing process of ribosomal precursor RNA in macronuclei of Tetrahymena thermophila. Nucleic Acids Res 15: 1905–1920.
7. Sorek R (2007) The birth of new exons: mechanisms and evolutionary consequences. Rna 13: 1603–1608.
8. Batzer MA, Deininger PL (2002) Alu repeats and human genomic diversity. Nat Rev Genet 3: 370–379.
9. Kriegs JO, Churakov G, Jurka J, Brosius J, Schmitz J (2007) Evolutionary history of 7SL RNA-derived SINEs in Supraprimates. Trends Genet 23: 158–161.
10. Cordaux R, Hedges DJ, Herke SW, Batzer MA (2006) Estimating the retrotransposition rate of human Alu elements. Gene 373: 134–137.
11. Deininger P, Batzer MA (1993) Evolution of retroposons. Evol Biol 27: 157–196.
12. Deininger PL, Batzer MA (1999) Alu repeats and human disease. Mol Genet Metab 67: 183–193.
13. Ng SK, Xue H (2006) Alu-associated enhancement of single nucleotide polymorphisms in the human genome. Gene 368: 110–116.
14. Makalowski W, Mitchell GA, Labuda D (1994) Alu sequences in the coding regions of mRNA: a source of protein variability. Trends Genet 10: 188–193.
15. Nekrutenko A, Li WH (2001) Transposable elements are found in a large number of human protein-coding genes. Trends Genet 17: 619–621.
16. Sorek R, Ast G, Graur D (2002) Alu-containing exons are alternatively spliced. Genome Res 12: 1060–1067.
17. Gal-Mark N, Schwartz S, Ram O, Eyras E, Ast G (2009) The pivotal roles of TIA proteins in 5′ splice-site selection of alu exons and across evolution. PLoS Genet 5: e1000717.
18. Gal-Mark N, Schwartz S, Ast G (2008) Alternative splicing of Alu exons—two arms are better than one. Nucleic Acids Res 36: 2012–2023.
19. Lev-Maor G, Sorek R, Shomron N, Ast G (2003) The birth of an alternatively spliced exon: 3′ splice-site selection in Alu exons. Science 300: 1288–1291.
20. Sorek R, Lev-Maor G, Reznik M, Dagan T, Belinky F, et al. (2004) Minimal conditions for exonization of intronic sequences: 5′ splice site formation in alu exons. Mol Cell 14: 221–231.
21. Corvelo A, Eyras E (2008) Exon creation and establishment in human genes. Genome Biol 9: R141.
22. Ram O, Schwartz S, Ast G (2008) Multifactorial interplay controls the splicing profile of Alu-derived exons. Mol Cell Biol 28: 3513–3525.
23. Schwartz S, Gal-Mark N, Kfir N, Oren R, Kim E, et al. (2009) Alu exonization events reveal features required for precise recognition of exons by the splicing machinery. PLoS Comput Biol 5: e1000300.
24. Krull M, Petrusma M, Makalowski W, Brosius J, Schmitz J (2007) Functional persistence of exonized mammalian-wide interspersed repeat elements (MIRs). Genome Res 17: 1139–1145.
25. Lin L, Jiang P, Shen S, Sato S, Davidson BL, et al. (2009) Large-scale analysis of exonized mammalian-wide interspersed repeats (MIRs) in primate genomes. Hum Mol Genet.
26. Lin L, Shen S, Tye A, Cai JJ, Jiang P, et al. (2008) Diverse splicing patterns of exonized alu elements in human tissues. PLoS Genet 4: e1000225.
27. Mersch B, Sela N, Ast G, Suhai S, Hotz-Wagenblatt A (2007) SERpredict: Detection of tissue- or tumor-specific isoforms generated through exonization of transposable elements. BMC Genet 8: 78.
28. Goren A, Kim E, Amit M, Bochner R, Lev-Maor G, et al. (2008) Alternative approach to a heavy weight problem. Genome Res 18: 214–220.
29. Kwan T, Benovoy D, Dias C, Gurd S, Provencher C, et al. (2008) Genome-wide analysis of transcript isoform variation in humans. Nat Genet 40: 225–231.
30. Graveley BR (2007) The haplo-spliceo-transcriptome: common variations in alternative splicing in the human population. Trends Genet.
31. Hull J, Campino S, Rowlands K, Chan MS, Copley RR, et al. (2007) Identification of common genetic variation that modulates alternative splicing. PLoS Genet 3: e99.
32. Kwan T, Benovoy D, Dias C, Gurd S, Serre D, et al. (2007) Heritability of alternative splicing in the human genome. Genome Res 17: 1210–1218.
33. Zhang XH, Chasin LA (2006) Comparison of multiple vertebrate genomes reveals the birth and evolution of human exons. Proc Natl Acad Sci U S A 103: 13427–13432.
34. Amit M, Sela N, Keren H, Melamed Z, Muler I, et al. (2007) Biased exonization of transposed elements in duplicated genes: A lesson from the TIF-IA gene. BMC Mol Biol 8: 109.
35. Ast G (2004) How did alternative splicing evolve? Nat Rev Genet 5: 773–782.
36. Hillman RT, Green RE, Brenner SE (2004) An unappreciated role for RNA surveillance. Genome Biol 5: R8.
37. Magen A, Ast G (2005) The importance of being divisible by three in alternative splicing. Nucleic Acids Res 33: 5574–5582.
38. Resch A, Xing Y, Alekseyenko A, Modrek B, Lee C (2004) Evidence for a subpopulation of conserved alternative splicing events under selection pressure for protein reading frame preservation. Nucleic Acids Res 32: 1261–1269.
39. Levy A, Sela N, Ast G (2008) TranspoGene and microTranspoGene: transposed elements influence on the transcriptome of seven vertebrates and invertebrates. Nucleic Acids Res 36: D47–52.
40. Roy M, Kim N, Xing Y, Lee C (2008) The effect of intron length on exon creation ratios during the evolution of mammalian genomes. Rna 14: 2261–2273.
41. Singer SS, Mannel DN, Hehlgans T, Brosius J, Schmitz J (2004) From "junk" to gene: curriculum vitae of a primate receptor isoform gene. J Mol Biol 341: 883–886.
42. Chang YF, Imam JS, Wilkinson MF (2007) The nonsense-mediated decay RNA surveillance pathway. Annu Rev Biochem 76: 51–74.
43. Lo WS, Lau CF, Xuan Z, Chan CF, Feng GY, et al. (2004) Association of SNPs and haplotypes in GABAA receptor beta2 gene with schizophrenia. Mol Psychiatry 9: 603–608.
44. Labuda D, Striker G (1989) Sequence conservation in Alu evolution. Nucleic Acids Res 17: 2477–2491.
45. Yang JO, Kim W-Y, Bhak J (2009) ssSNPTarget: genome-wide splice-site single nucleotide polymorphism database. Human Mutation 30: E1010–E1020.
46. Kent WJ, Sugnet CW, Furey TS, Roskin KM, Pringle TH, et al. (2002) The human genome browser at UCSC. Genome Res 12: 996–1006.
47. Lev-Maor G, Sorek R, Levanon EY, Paz N, Eisenberg E, et al. (2007) RNA-editing-mediated exon evolution. Genome Biol 8: R29.

48. Moller-Krull M, Zemann A, Roos C, Brosius J, Schmitz J (2008) Beyond DNA: RNA editing and steps toward Alu exonization in primates. J Mol Biol 382: 601–609.
49. Athanasiadis A, Rich A, Maas S (2004) Widespread A-to-I RNA editing of Alu-containing mRNAs in the human transcriptome. PLoS Biol 2: e391.
50. Blow M, Futreal PA, Wooster R, Stratton MR (2004) A survey of RNA editing in human brain. Genome Res 14: 2379–2387.
51. Kim DD, Kim TT, Walsh T, Kobayashi Y, Matise TC, et al. (2004) Widespread RNA editing of embedded alu elements in the human transcriptome. Genome Res 14: 1719–1725.
52. Levanon EY, Eisenberg E, Yelin R, Nemzer S, Hallegger M, et al. (2004) Systematic identification of abundant A-to-I editing sites in the human transcriptome. Nat Biotechnol 22: 1001–1005.
53. He S, Liu C, Skogerbo G, Zhao H, Wang J, et al. (2008) NONCODE v2.0: decoding the non-coding. Nucleic Acids Res 36: D170–172.
54. Liu C, Bai B, Skogerbo G, Cai L, Deng W, et al. (2005) NONCODE: an integrated knowledge database of non-coding RNAs. Nucleic Acids Res 33: D112–115.
55. Nakabayashi K, Bentley L, Hitchins MP, Mitsuya K, Meguro M, et al. (2002) Identification and characterization of an imprinted antisense RNA (MESTIT1) in the human MEST locus on chromosome 7q32. Hum Mol Genet 11: 1743–1756.
56. Nemes JP, Benzow KA, Moseley ML, Ranum LP, Koob MD (2000) The SCA8 transcript is an antisense RNA to a brain-specific transcript encoding a novel actin-binding protein (KLHL1). Hum Mol Genet 9: 1543–1551.
57. Chen WL, Lin JW, Huang HJ, Wang SM, Su MT, et al. (2008) SCA8 mRNA expression suggests an antisense regulation of KLHL1 and correlates to SCA8 pathology. Brain Res 1233: 176–184.
58. Koob MD, Moseley ML, Schut LJ, Benzow KA, Bird TD, et al. (1999) An untranslated CTG expansion causes a novel form of spinocerebellar ataxia (SCA8). Nat Genet 21: 379–384.
59. Xing Y, Lee C (2006) Alternative splicing and RNA selection pressure—evolutionary consequences for eukaryotic genomes. Nat Rev Genet 7: 499–509.
60. Gerber A, O'Connell MA, Keller W (1997) Two forms of human double-stranded RNA-specific editase 1 (hRED1) generated by the insertion of an Alu cassette. Rna 3: 453–463.
61. Gehring NH, Lamprinaki S, Hentze MW, Kulozik AE (2009) The hierarchy of exon-junction complex assembly by the spliceosome explains key features of mammalian nonsense-mediated mRNA decay. PLoS Biol 7: e1000120.
62. Kim E, Magen A, Ast G (2007) Different levels of alternative splicing among eukaryotes. Nucleic Acids Res 35: 125–131.
63. Xie H, Wang M, Bonaldo MD, Smith C, Rajaram V, et al. (2009) High-throughput sequence-based epigenomic analysis of Alu repeats in human cerebellum. Nucleic Acids Res.
64. Mills RE, Bennett EA, Iskow RC, Devine SE (2007) Which transposable elements are active in the human genome? Trends Genet 23: 183–191.
65. Gibbs RA, Rogers J, Katze MG, Bumgarner R, Weinstock GM, et al. (2007) Evolutionary and biomedical insights from the rhesus macaque genome. Science 316: 222–234.
66. Kuhn RM, Karolchik D, Zweig AS, Trumbower H, Thomas DJ, et al. (2007) The UCSC genome browser database: update 2007. Nucleic Acids Res 35: D668–673.
67. Karolchik D, Hinrichs AS, Furey TS, Roskin KM, Sugnet CW, et al. (2004) The UCSC Table Browser data retrieval tool. Nucleic Acids Res 32: D493–496.
68. Giardine B, Riemer C, Hardison RC, Burhans R, Elnitski L, et al. (2005) Galaxy: a platform for interactive large-scale genome analysis. Genome Res 15: 1451–1455.
69. Jurka J (2000) Repbase update: a database and an electronic journal of repetitive elements. Trends Genet 16: 418–420.
70. Smit AFA, Hubley R, Green P. Version: open-3.2.7.
71. Jurka J, Kapitonov VV, Pavlicek A, Klonowski P, Kohany O, et al. (2005) Repbase Update, a database of eukaryotic repetitive elements. Cytogenet Genome Res 110: 462–467.
72. Kapitonov VV, Jurka J (2008) A universal classification of eukaryotic transposable elements implemented in Repbase. Nat Rev Genet 9: 411–412; author reply 414.
73. Sherry ST, Ward MH, Kholodov M, Baker J, Phan L, et al. (2001) dbSNP: the NCBI database of genetic variation. Nucleic Acids Res 29: 308–311.